\documentclass{PoS}

\title{Prompt atmospheric neutrino flux}

\ShortTitle{Amospheric neutrinos}

\author{\speaker{Yu Seon Jeong} \\
        National Institute of Supercomputing and Networking, KISTI, Daejeon 34141, Korea\\
        E-mail: \email{ysjeong@kisti.re.kr}}

\author{Atri Bhattacharya\\
            Space sciences, Technologies and Astrophysics Research (STAR) Institute,
                Universit\'{e} de Li\`{e}ge, B\^{a}t.~B5a, 4000 Li\`{e}ge,
                Belgium\\
        E-mail: \email{a.bhattacharya@ulg.ac.be}}

\author{Rikard Enberg\\
        Department of Physics and Astronomy, Uppsala University, Uppsala, Sweden\\
        E-mail: \email{rikard.enberg@physics.uu.se}}
        
\author{C. S. Kim\\
	Department of Physics and IPAP, Yonsei University, Seoul 03722, Korea\\
        E-mail: \email{cskim@yonsei.ac.kr}}
        
\author{Mary Hall Reno\\
	Department of Physics and Astronomy, University of Iowa, Iowa City, Iowa 52242\\	
        E-mail: \email{mary-hall-reno@uiowa.edu}}
        
\author{Ina Sarcevic\\
	Department of Astronomy, University  of Arizona, 933 N.\ Cherry Ave., Tucson, AZ 85721\\
        E-mail: \email{ina@physics.arizona.edu}}

\author{Anna Stasto\\
	Department of Physics, The Pennsylvania State University, University Park, PA 16802\\
        E-mail: \email{ams52@psu.edu}}

\abstract{We evaluate the prompt atmospheric neutrino flux including nuclear correction and $B$ hadron contribution
in the different frameworks: NLO perturbative QCD and dipole models. 
The nuclear effect is larger in the prompt neutrino flux than in the total charm production cross section, 
and it reduces the fluxes by $10\% - 30\%$ depending on the model.
We also investigate the uncertainty using the QCD scales allowed by the charm  cross section data 
from RHIC and LHC experiments.    
}

\FullConference{38th International Conference on High Energy Physics\\
		3-10 August 2016\\
		Chicago, USA}

\begin{document}

\section{Introduction}
\noindent
Numerous cosmic ray particles are incident into the Earth and collide with the nuclei in the upper atmosphere. 
From the interactions of cosmic rays with the nuclei, various kinds of hadrons are produced, 
and some of them subsequently decay generating neutrinos.
Neutrinos produced in this process are called atmospheric neutrinos.
There are two components in atmospheric neutrinos: conventional neutrinos from the  decays of $\pi$ or $K$ mesons
and prompt neutrinos from the decays of heavy hadrons.
Although conventional neutrinos are typical atmospheric neutrinos, its flux rapidly decreases with energy
due to the energy loss of $\pi$ or $K$ through interactions before they decay.  
On the other hand, the heavy hadrons immediately decay, even at high energies, due to their extremely short lifetimes.
As a result, the flux of prompt neutrinos dominates at high energy ($E \gtrsim$ 1 PeV). 

Atmospheric neutrinos are an important background to the astrophysical neutrinos. 
At present, the IceCube experiment has detected 54 high energy neutrino candidate events 
between the energies of 20 TeV and 2 PeV, and excluded a pure atmospheric explanation for their origin with high accuracy \cite{icecube}.
The prompt neutrinos can be the primary background to the events at such high energies,
hence it is necessary to estimate this flux precisely for the analysis of the experimental measurements.  

Among earlier evaluations of the prompt neutrino flux, the prediction by some of us (called ERS) \cite{ers} has been used 
as the benchmark for the analysis of the IceCube measurement. 
The ERS flux was evaluated based on the so-called dipole model for the heavy quark production.  
Recently, they re-evaluated the prompt flux in the next-to-leading order perturbative QCD (NLO pQCD) 
using the updated parton distribution function (PDF), CT10 
and more elaborated cosmic ray spectra in \cite{berss} (BERSS). The BERSS flux reduces the ERS flux by a factor of 1.5. 

In this proceeding paper, we present the improved prompt neutrino fluxes.
In particular, we include nuclear corrections and the $B$ hadron contributions 
with the recent PDF sets and charm fragmentation fractions.
We use the experimentally constrained QCD scales to investigate the uncertainty in the prompt neutrino fluxes.   
Our comprehensive works with more detailed information are presented in ref. \cite{bejkrss}. 
Other recent evaluations appear in \cite{others}.

\section {Calculation of the atmospheric neutrino flux}
\noindent
The atmospheric neutrino flux can be calculated using the cascade equations, 
which can be solved by the semi analytic $Z$-moment method. 
The cascade equations describe the propagation of particles in the atmosphere in terms of the change of their fluxes. 
Its general form is given by 
\begin{equation}
\frac{d \phi_j}{d X} = -\frac{\phi_j}{\lambda_j} - \frac{\phi_j}{\lambda^{dec}_j} + \sum_k S(k \rightarrow j) 
\label{eq:cascade}
\end{equation}
with the interaction (decay) length $\lambda_j^{(dec)}$ and the generation function $S(k \rightarrow j)$.
Here, the slant depth $X$ represents the amount of the atmosphere traversed by a particle.  
The generation function, which depends both on the energy ($E$) and the slant depth ($X$), 
can be rescaled to have only energy dependence.
This is called $Z$-moment, and for particle production, it can be written as 
\begin{eqnarray}
Z_{kj}(E)  \simeq  S(k\rightarrow j) \frac{\lambda_k(E)} {\phi_k(E,X)} 
 = \int ^{\infty}_{E} dE ' \frac{\phi_k(E')} {\phi_k(E)} \frac{\lambda_k(E)} {\lambda_k(E')}
\frac{1}{\sigma_{kA}(E')}\frac{d\sigma(kA\to jY;E',E)}{dE } \, 
\label{eq:fgen}
\end{eqnarray}
with the assumption, $\phi_k(E,X) \cong \phi_k(E) f(X)$.  
Then, the solutions of a set of the coupled cascade equations for protons, hadrons and neutrinos 
can be written in terms of the $Z$-moments for production of hadrons and those for the decay to neutrinos
in the high energy and low energy limits.
The final neutrino flux is evaluated by interpolating these two approximate solutions and doing the sum over hadrons. 
The $Z$-moment method for evaluating the flux is described in details in ref. \cite{lipari}.
In our evaluation, $D^0$, $D^+$, $D_s$, $\Lambda_c$ ($B^0$, $B^+$, $B_s$, $\Lambda_b$) were included for the charmed (bottom) hadrons.

\section{Heavy quark production cross sections}

\subsection{Charm/bottom production in perturbative QCD}
\noindent
The essential ingredient for calculating the prompt flux is the production cross sections for heavy quarks,
which creates the charmed/bottom hadrons through fragmentation. 
First we compute the heavy quark production cross section in the perturbative QCD at NLO. 
Nuclear corrections are incorporated by the recent nuclear PDF, nCTEQ15 \cite{ncteq15}. 
For the factorization ($M_F$) and renormalization ($M_R$) scales, we take $(2.1, 1.6) m_q$ for central values, 
experimentally constrained by the RHIC and LHC data \cite{nvf}. 

\begin{figure}
\centering
 \includegraphics[width=7.5cm]{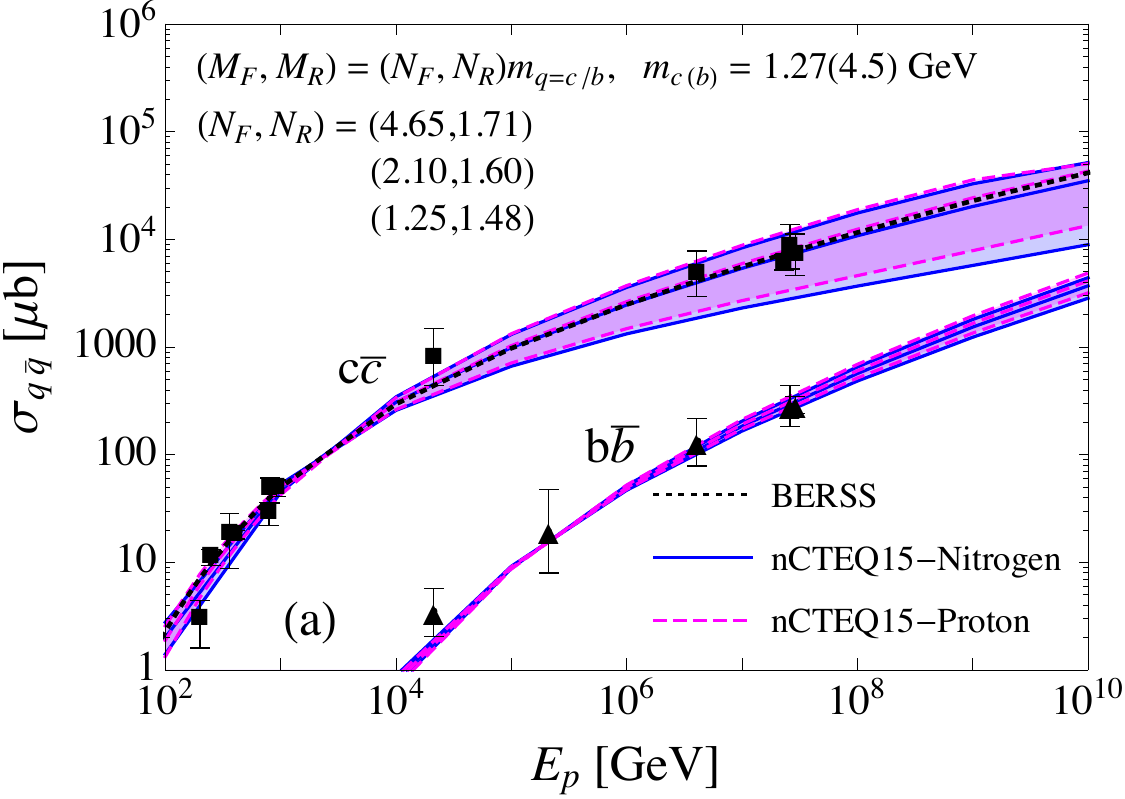}
  \includegraphics[width=7.5cm]{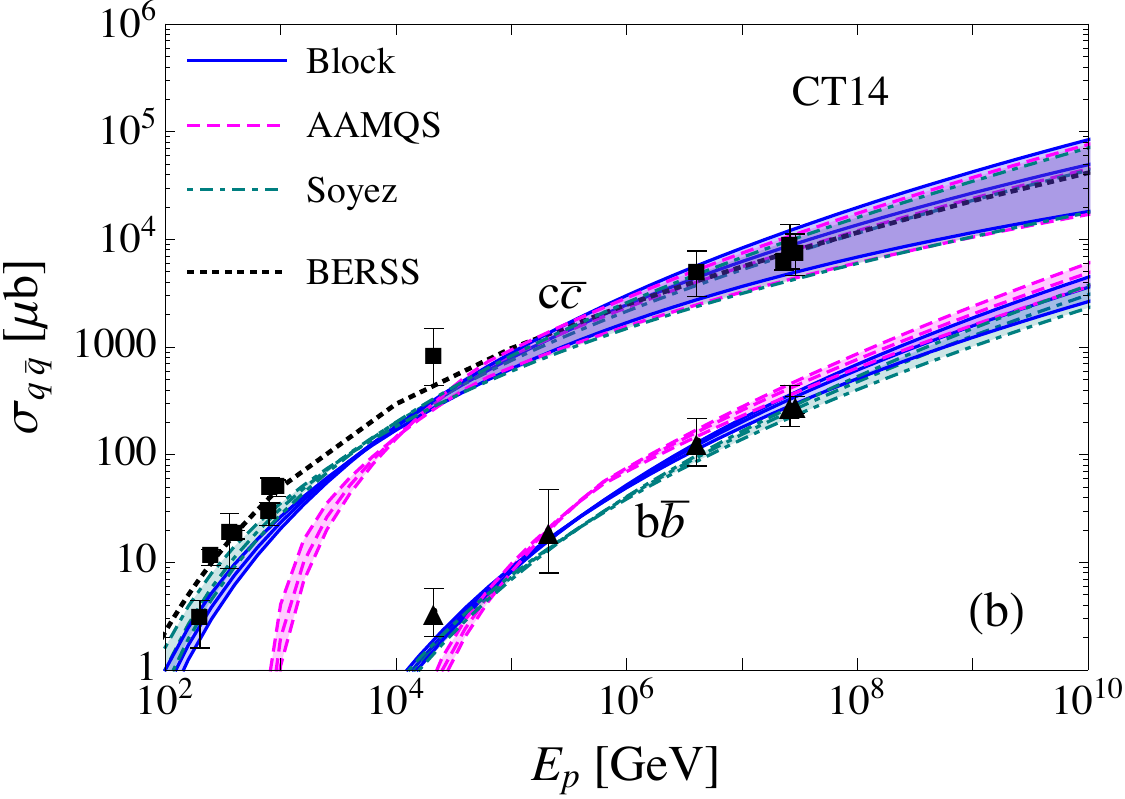}
\caption{ Total cross section for $c\bar{c}$ and $b\bar{b}$ production as a function of the incident proton energy 
evaluated (a) in the perturbative QCD using the nCTEQ15 PDFs and (b) in the dipole model using the CT14 LO PDF.
For comparison the previous evaluation at NLO in pQCD approach is presented as BERSS \cite{berss}.}
\label{fig-hqx}     
\end{figure}

In Fig. \ref{fig-hqx} (a), we show the total cross sections per nucleon in $pp$ and $p-nitrogen$ collisions for the charm and bottom  production.
The error bands indicate the uncertainty in the cross sections from the scale variation $(1.25, 1.48) m_q$ and $(4.65, 1.71) m_q$ for the lower and upper limits, suggested in ref. \cite{nvf}. 
We use this scale range to evaluate the neutrino flux as well.
The data points come from the RHIC, LHC and fixed target experiments \cite{xsection-data, lhcb}.

The impact of the nuclear correction on the cross section is presented in fig. \ref{fig-nucX}  (a)
with the ratio of the cross sections per nucleon for nitrogen and free nucleons. 
We find that the nuclear effect increases with energy 
and the cross section is reduced by $\sim$ 20\% (10\%) at $E=10^{10}$ GeV for the charm (bottom) production.

\subsection{Dipole model}
\noindent
Another possible approach is a dipole model, 
in which an incoming gluon splits into $q \bar{q}$ pair and interacts with target nucleus as a dipole.  
The cross section for the heavy quark production in the dipole model can be written as 
\begin{equation}
\frac{d\sigma (pp\to q\bar{q}X)}{dx_F}\simeq \frac{x_1}{x_1+x_2}
 g(x_1,M_F)\sigma^{gp\to q\bar{q}X}
(x_2,M_R,Q^2=0)\ ,
\end{equation}
where the partonic cross section $\sigma^{gp\to q\bar{q}X}$ is
\begin{equation}
\sigma^{g p \to q\bar q X}(x,M_R,Q^2) = \int dz \, d^2\vec r \,
|\Psi^q_g(z,\vec r,M_R,Q^2)|^2 \sigma_{d}(x,\vec r)\ ,
\label{eq:dipolexsec}
\end{equation}
with the wave function squared for the gluon fluctuation,
and the dipole cross section ($\sigma_d$) for the interaction of the dipole with the target. 

We employ the Glauber-Gribov formalism to include the nuclear correction, given by
\begin{equation}
\sigma_d^A(x,r)=\int d^2\vec{b}\, 2\Biggl[ 1-\exp\Biggl( -\frac{1}{2} AT_A(b)\sigma_d^p(x,r)\Biggr)\Biggr] \ , 
\end{equation}
with the impact parameter $b$. 
The nuclear profile function $T_A(b) = \int d z \rho_A (z, \vec{b})$ is normalized to unity, $\int d^2 \vec{b}\, T_A(b)= 1$.  
Here, we use a Gaussian distribution for nuclear density, $\rho_A (z, \vec{b})$.

Fig. \ref{fig-hqx} (b) shows the total cross sections in $pp$ collisions for the charm and bottom production 
evaluated using three different different dipole cross sections.
The Soyez dipole cross section \cite{soyez} was used for the prediction of the ERS flux in ref. \cite{ers}. 
Other dipoles referred to as 'AAMQS' \cite{aamqs} and 'Block'\cite{blockdm} are more recent and/or improved dipole cross sections.
(For the details, see the corresponding references or brief description in ref. \cite{bejkrss}.)
The range of the cross sections are due to the factorization scale in the gluon PDF.
Comparing the experimental data, we found that the allowed scale range is $m_c \leq M_F \leq 4 m_c$.

\begin{figure}
\centering
\includegraphics[width=7.5cm,clip]{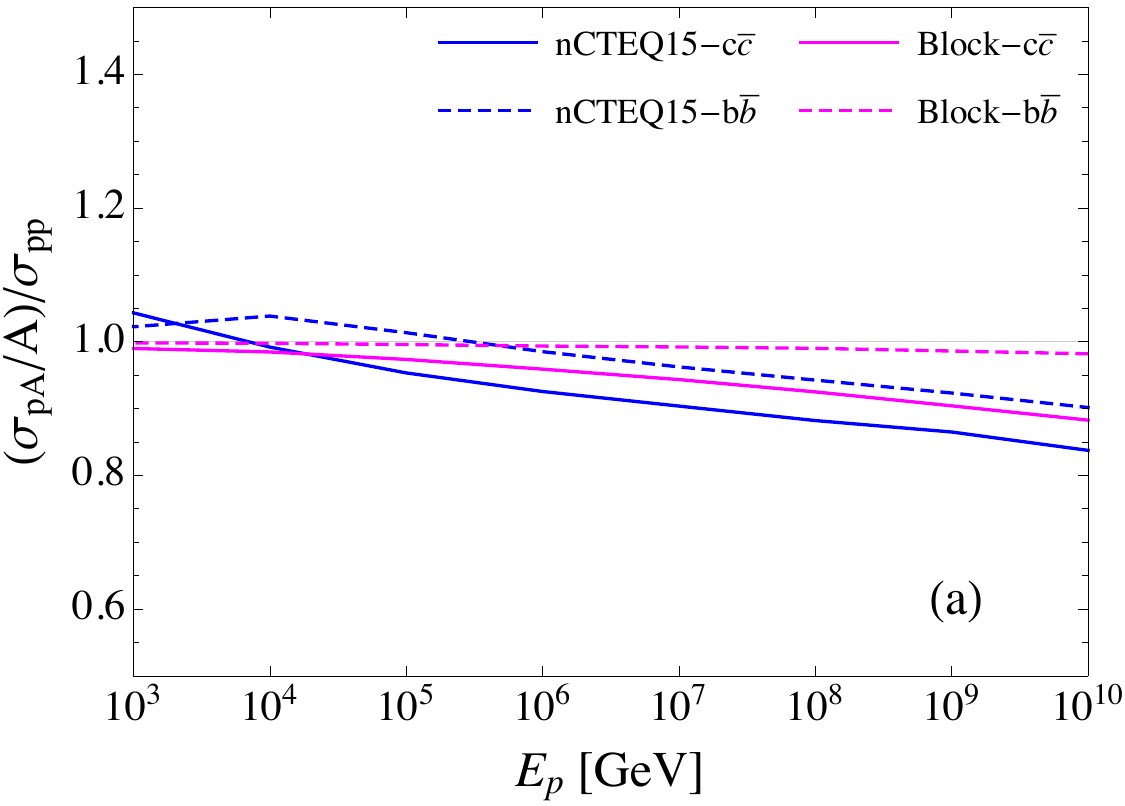}
\includegraphics[width=7.5cm,clip]{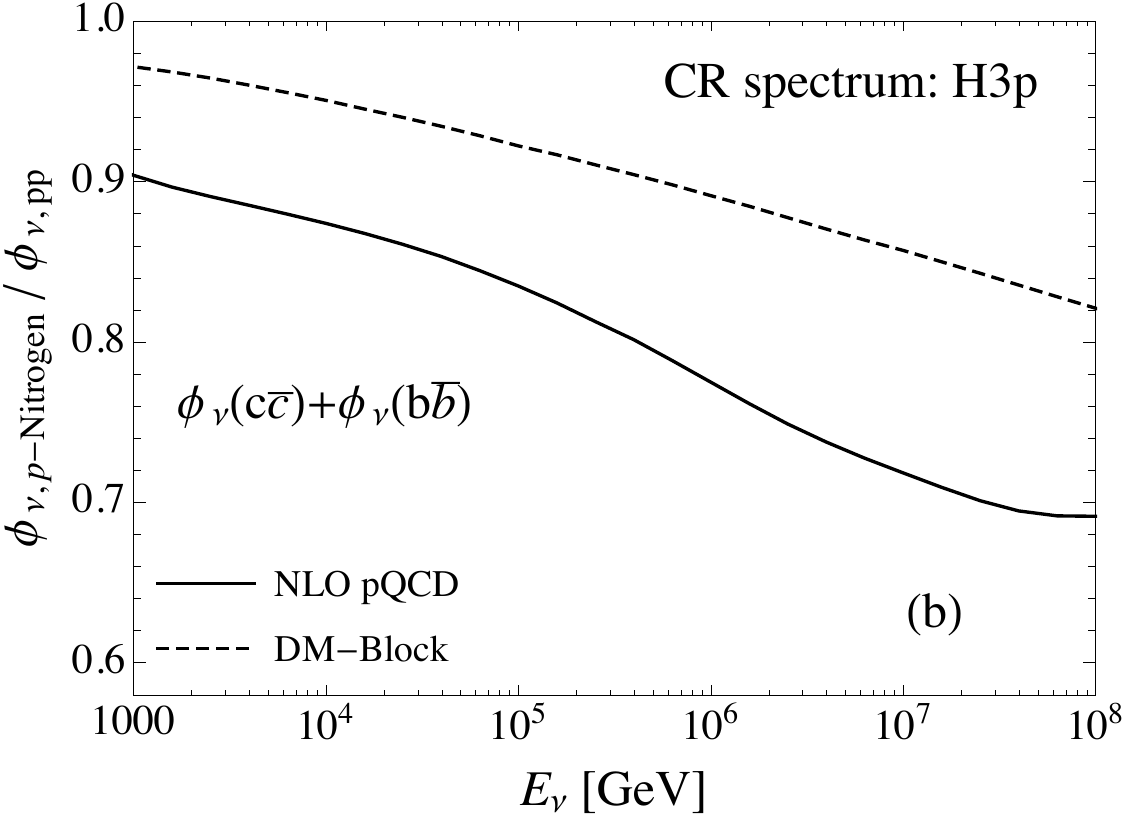}
\caption{Nuclear effects in the total cross sections for the $q\bar{q}$ production (a) and in the prompt muon neutrino fluxes (b) }
\label{fig-nucX}     
\end{figure}

Nuclear effects in the dipole models are smaller than in the perturbative QCD, which are about
10 \% (2 \%) at $10^{10}$ GeV for the charm (bottom) production.
In fig. \ref{fig-nucX} (a), we present the results only for the Block dipole to avoid cluttering. 

\section{Prompt atmospheric neutrino fluxes}
\subsection{Cosmic ray fluxes}
\noindent
In evaluating the prompt neutrino fluxes, we use three cosmic ray fluxes as follows: 
(a) Broken power law (BPL) $-$ parameterized as
$\phi_N(E)  = 1.7\ (E/{\rm GeV})^{-2.7}$ for $ E< 5\cdot 10^6$ GeV 
and $174 \ (E/{\rm GeV})^{-3}$ for $E>5\cdot 10^6$ GeV 
assuming cosmic rays consist only of protons.
(b) H3p $-$ parameterized based on the model for three source components 
(SN remnants, galactic sources and extragalactic sources) considering the compositions of cosmic rays \cite{gaisser}.
Cosmic rays from extragalactic sources are all protons.
(c) H3a $-$ similar to H3p, except the mixed composition from the extragalactic sources.  

\subsection{Prompt muon neutrino flux}
\begin{figure}
\centering
\includegraphics[width=7.5cm,clip]{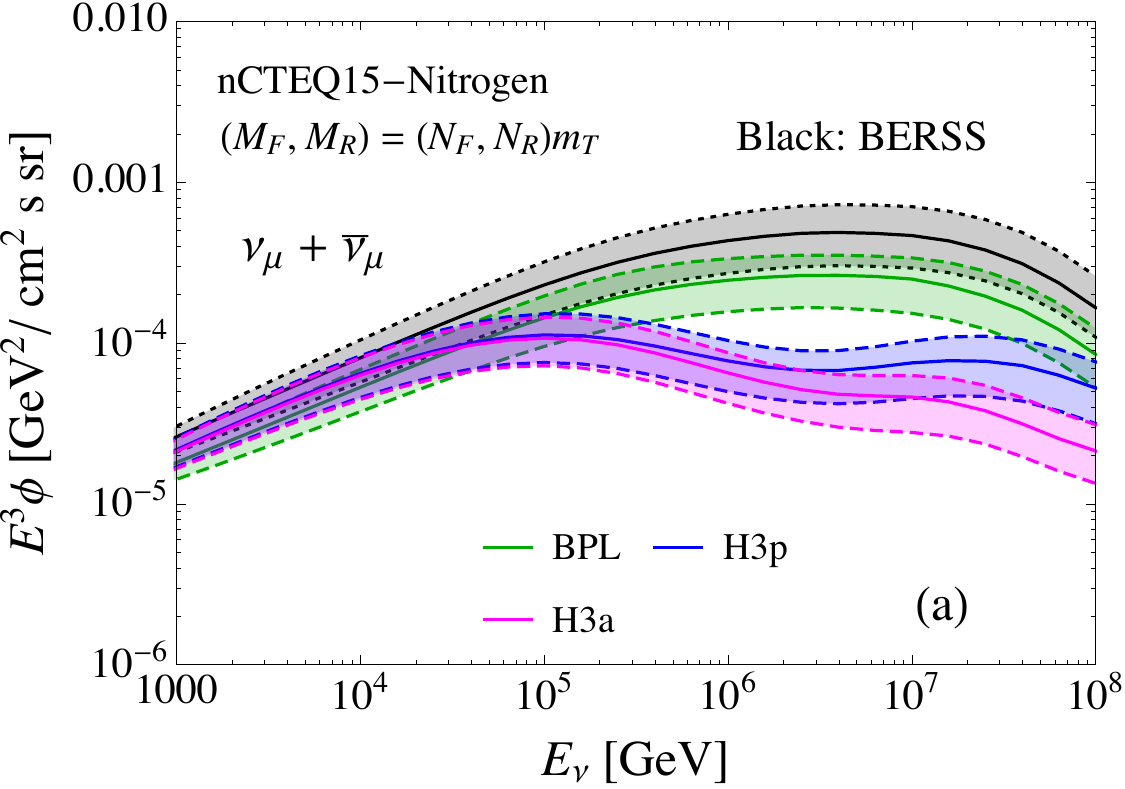}
\includegraphics[width=7.5cm,clip]{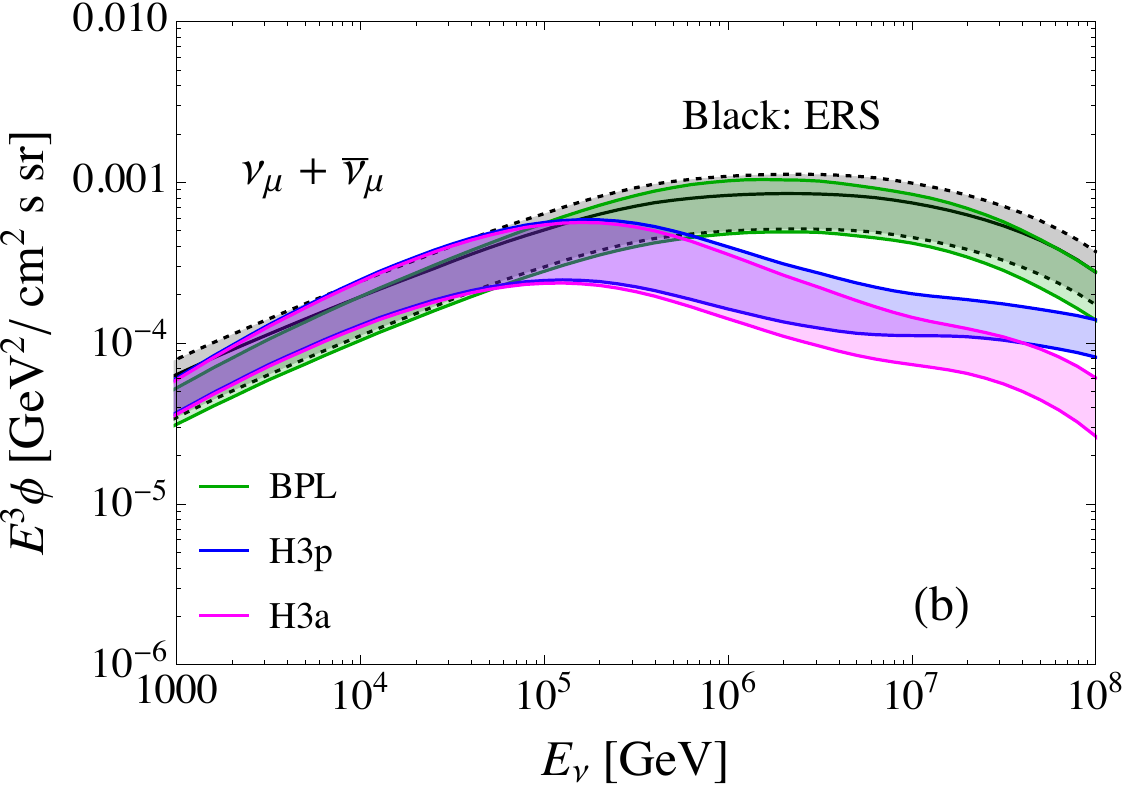}
\caption{The prompt atmospheric neutrino fluxes evaluated with nuclear correction (a) in the pQCD at NLO and (b) in the dipole models.
The BERSS and ERS fluxes are also shown for comparison.}
\label{fig-flux}     
\end{figure}

\noindent
Fig. \ref{fig-flux} shows our resulting neutrino fluxes evaluated with nuclear corrections in the NLO pQCD and the dipole models. 
For comparison, we present the BERSS and ERS fluxes for the BPL in the pQCD and dipole model results, respectively.  
For the pQCD case, the error bands are due to the scale variation experimentally constrained by the cross section data as discussed above.
Differences from the BERSS flux are due to nuclear corrections, $B$ hadron contributions, the new PDF set and new fragmentation fraction for charm.
The combined effect of these results in our new fluxes being $\sim 30\%-45\%$ lower than the BERSS.
The largest impact is from the nuclear corrections, which reduces the flux about $20\%-30\%$ at $E_\nu = 10^5-10^8$ GeV
as shown in fig. \ref{fig-nucX} (b).

For the fluxes from the dipole models, the error bands reflect the uncertainty due to the various dipole cross sections 
as well as the scale variation.  
In addition to the updated factors in the pQCD evaluations, the dipole model results have improved 
with some updated $Z$ moments, $Z_{pp}$ and $Z_{DD}$, and the charm mass.  
The nuclear effect in the dipole model evaluations is a $10\%-20\%$ reduction at $10^5-10^8$ GeV.
However, as we can see in fig. \ref{fig-flux} (b), the fluxes with all the updates are comparable to the ERS fluxes.

\section {Summary and discussion}
\noindent
We have evaluated the nuclear corrected prompt neutrino flux in the different frameworks, 
NLO pQCD and the dipole models. 
We also used the QCD scales constrained by the charm cross section data from RHIC and LHC to investigate the uncertainty. 
The nuclear effect is different according to the models, which is larger in the predictions from NLO pQCD than from the dipole models.
The nuclear correction has a more significant impact on the prompt flux, where the forward region is more important, than on the total charm production cross section.

The new prompt neutrino fluxes from NLO pQCD are lower than those from the dipole models,
and they do not violate the IceCube limit based on the 3 year observation \cite{icecube}, which excludes most of our dipole model predictions.    
However, we note that even the fluxes from the dipole models are below the new limit from the 6 year data \cite{icecube-six}, which was released after our work.

\section* {Acknowledgments}
This research was supported by the US DOE, the NRF of Korea, the NSC of Poland, 
the SRC of Sweden and the FNRS of Belgium.

\end{document}